\RequirePackage{lineno}
\documentclass[twocolumn,aps,prc,showpacs,superscriptaddress,preprintnumbers,floatfix,nofootinbib]{revtex4}
\usepackage{epsfig,graphics}
\usepackage{graphicx}
\usepackage{dcolumn}
\usepackage{bm}
\usepackage{amsmath}
\usepackage[usenames]{color}
\usepackage{ulem} 

\begin{document}

\title{Azimuthal anisotropies of reconstructed jets in Pb+Pb collisions at $\sqrt{s_{_{\rm NN}}}$ = 2.76 TeV in a multiphase transport model}

\author{Mao-Wu Nie}
\affiliation{Shanghai Institute of Applied Physics, Chinese Academy of Sciences, Shanghai 201800, China}
\affiliation{University of Chinese Academy of Sciences, Beijing 100049, China}

\author{Guo-Liang Ma}
\affiliation{Shanghai Institute of Applied Physics, Chinese Academy of Sciences, Shanghai 201800, China}


\begin{abstract}
Azimuthal anisotropies of reconstructed jets [$v_{n}^{jet} (n=2, 3)$] have been investigated in Pb+Pb collisions at the center of mass energy $\sqrt{s_{_{\rm NN}}}$ = 2.76 TeV within a framework of a multiphase transport (AMPT) model. The $v_{2}^{jet}$ is in good agreement with the recent ATLAS data. However, the $v_{3}^{jet}$ shows a smaller magnitude than $v_{2}^{jet}$, and approaches zero at a larger transverse momentum. It is attributed to the path-length dependence in which the jet energy loss fraction depends on the azimuthal angles with respect to different orders of event planes. The ratio $v_{n}^{jet}/\varepsilon_{n}$ increases from peripheral to noncentral collisions, and $v_{n}^{jet}$ increases with the initial spatial asymmetry ($\varepsilon_{n}$) for a given centrality bin. These behaviors indicate that the $v_{n}^{jet}$ is produced by the strong interactions between jet and the partonic medium with different initial geometry shapes. Therefore, azimuthal anisotropies of reconstructed jet are proposed as a good probe to study the initial spatial fluctuations, which are expected to provide constraints on the path-length dependence of jet quenching models.
\end{abstract}


\maketitle

\section{Introduction}
\label{sec:intro}
A deconfined quark-gluon plasma (QGP) could be created in the early state of high-energy heavy-ion collisions at the Relativistic Heavy Ion Collider (RHIC) and the CERN Large Hadron Collider (LHC)~\cite{oai:arXiv.org:nucl-ex/0501009, oai:arXiv.org:nucl-ex/0410003}. A jet, produced by initial hard processes, is an important probe to understand the properties of the QGP, since it losses its energy when it passes through the hot partonic medium~\cite{Wang:1991xy}. This phenomenon, so-called jet quenching, has been confirmed by many experimental observations. For example, the nuclear modification factor $R_{AA}$ shows a strong suppression at high transverse momentum $p_{T}$ in central A+A collisions at RHIC~\cite{oai:arXiv.org:nucl-ex/0305015} and LHC~\cite{oai:arXiv.org:1012.1004} energies. The measured elliptic anisotropy (or elliptic ``flow") $v_{2}$ of final hadrons remains positive above $\sim$10 GeV/$c$ in A+A collisions at the RHIC~\cite{oai:arXiv.org:1006.3740} and LHC~\cite{oai:arXiv.org:1204.1409} energies, which discloses a path-length dependence of jet quenching~\cite{oai:arXiv.org:nucl-th/0012092}. Besides these above jet measurements based on high-$p_T$ leading hadrons, the recent LHC measurements on fully reconstructed jets provide a comprehensive characterization of jet quenching. For instance, a larger dijet $p_{T}$ asymmetry has been observed in central Pb+Pb collisions than in p+p collisions at the LHC energy~\cite{oai:arXiv.org:1011.6182,oai:arXiv.org:1102.1957}, which is thought to be additional direct evidence of jet energy loss in the QGP, as important as the disappearance of the away-side peak in dihadron azimuthal correlation in central Au+Au collisions at the top RHIC energy~\cite{oai:arXiv.org:nucl-ex/0210033}. The data on the elliptic anisotropy of reconstructed jets are recently released by the ATLAS Collaboration, which show nonzero $v_{2}$ values for the $p_{T}$ range from 45 to 160 GeV/$c$ for all centrality bins in Pb+Pb collisions~\cite{Aad:2013sla,ATLAS:2012hna}. It is consistent with a path-length dependence of jet energy loss with respect to the reaction plane. The elliptic anisotropy of reconstructed jets can be theoretically reproduced by the JEWEL model within a perturbative framework for jet evolution in a QGP medium~\cite{Zapp:2013zya}. The recent studies of higher orders of harmonic flow, especially for triangular flow $v_{3}$, have deepened our understanding of many aspects of high energy heavy-ion collisions~\cite{Alver:2010gr, Schenke:2010rr, Lacey:2010yg}. It would be interesting to study the third order of anisotropy $v_{3}$ of reconstructed jets, as it serves as the jet response to the initial geometry triangularity which could provide a greater constraint on jet quenching models. The conversion efficiency $v_{n}^{jet}/\varepsilon_{n}$, the ratio of jet $v_{n}$ over the initial spatial eccentricity, is also an important observable to learn about how the energy loss of reconstructed jets depends the initial geometry asymmetry. As heavy-ion collisions are dynamical evolutions, it is also necessary for understanding the whole jet quenching picture to study how reconstructed jets evolve dynamically during different evolution stages. 

In this work, the elliptic anisotropy $v_{2}$ and triangular anisotropy $v_{3}$ of reconstructed jets are investigated in Pb+Pb collisions at $\sqrt{s_{_{\rm NN}}}$ = 2.76 TeV within a multiphase transport (AMPT) model, which includes both dynamical evolutions of partonic and hadronic phases. In the remaining part of this paper, we refer to ``jet'' as a fully reconstructed jet for simplicity unless otherwise noted. We find that the AMPT model can well describe the experimental results about jet $v_{2}$. Jet $v_{n}$ (n=2 and 3) arises owing to the strong interactions between jets and a partonic matter with different geometrical asymmetries. The other final state interactions, such as hadronization (via coalescence) and hadronic rescatterings, show little impact on the measured jet $v_{n}$.  We do observe that the jet energy loss fraction is dependent on the azimuthal angle with respect to the different orders of event plane. We find that jet $v_{n}$ is sensitive to the spatial eccentricity ($\varepsilon_{n}$) of initial parton distribution. We further propose azimuthal anisotropies of reconstructed jets as a good probe to study the initial spatial fluctuations, and expect  that jet $v_{n}$ provides constraints on the path-length dependence of jet quenching models.

\section{The AMPT Model}
\label{sec:model}
The AMPT model with the string melting mechanism is utilized in this work~\cite{Lin:2004en}. It consists of four main stages of high-energy heavy-ion collisions: the initial condition, parton cascade, hadronization, and hadronic rescatterings. In order to increase the simulation efficiency of jets with $p_{T} > 45$ GeV/$c$, a dijet of $p_{T}\sim$ 40 GeV/$c$ is triggered in the initial condition based on the HIJING model~\cite{Wang:1991hta,Gyulassy:1994ew}. The high-$p_{T}$ primary partons evolve into jet showers full of lower virtuality partons through initial- and final- state QCD radiations. In the string melting mechanism, all excited strings and jets are fragmented into hadrons according to the Lund string fragmentation~\cite{Sjostrand:2000wi}. Then these hadrons are converted to quarks according to the flavor and spin structures of their valence quarks. After the melting process, the jet parton showers are converted into clusters of on-shell constituent quarks and anti-quarks, and a plasma of on-shell constituent quarks and anti-quarks is also formed. Next, Zhang's parton cascade (ZPC) model~\cite{Zhang:1997ej} automatically simulates all possible elastic partonic interactions among the medium quarks and jet shower quarks, but without including inelastic parton interactions or further radiations at present. When the quarks freeze out, they are recombined into medium hadrons or jet shower hadrons via a simple coalescence model which combines two nearest quarks into a meson and three nearest quarks into a baryon. The final-state hadronic interactions, including elastic and inelastic hadronic scatterings and resonance decays, can be described by a relativistic hadronic transport (ART) model~\cite{Li:1995pra}. For more details on the AMPT model, we refer the reader to Ref.~\cite{Lin:2004en}. Recently, the AMPT model with a partonic interaction cross section of 1.5 mb has successfully given many qualitative descriptions of the experimental results about pseudorapidity and $p_{T}$ distributions~\cite{Xu:2011fi}, harmonic flows~\cite{Xu:2011fe, Xu:2011jm}, and reconstructed jet observables, including $\gamma$-jet $p_{T}$ imbalance~\cite{Ma:2013bia}, dijet $p_{T}$ asymmetry~\cite{Ma:2013pha}, jet fragmentation function~\cite{Ma:2013gga} and jet shape~\cite{Ma:2013uqa} in Pb+Pb collisions at $\sqrt{s_{_{\rm NN}}}$ = 2.76 TeV. Consistently with the previous studies, a partonic interaction cross section, 1.5 mb, is kept to simulate Pb+Pb collisions at $\sqrt{s_{_{\rm NN}}}$ = 2.76 TeV in this work.

\section{Jet Reconstruction}
\label{sec:jetrec}
To fully reconstruct jets, our kinematic cuts are chosen to be the same as in the ATLAS experiment~\cite{Aad:2013sla,ATLAS:2012hna}. An anti-$k_{t}$ algorithm from the standard Fastjet package is used to reconstruct full jets~\cite{Cacciari:2011ma}, in which the jet cone size $R$ is set to be 0.2. A pseudorapidity strip of width $\Delta\eta$=1.0 centered on the jet position, with two highest-energy jets excluded, is used to estimate the background (``average energy per jet area"), which is subtracted from the reconstructed jet energy in Pb+Pb collisions. Only jets within a mid-rapidity range of $|\eta|<2$ are considered in our analysis.

\section{Results and Discussions}
\label{sec:results}

\begin{figure}
\includegraphics[scale=0.43]{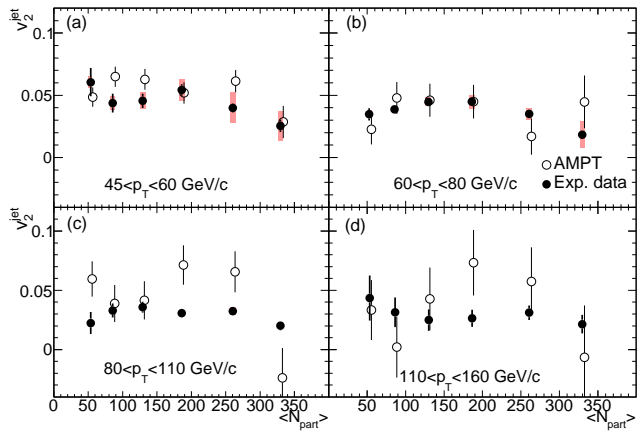}
\caption{(Color online) $v_{2}^{jet}$ as functions of $N_{part}$ for four different jet $p_{T}$ bins in Pb+Pb collisions at $\sqrt{s_{_{\rm NN}}}$ = 2.76 TeV, where open circles represent the AMPT results and solid circles represent the ATLAS experimental data~\cite{Aad:2013sla,ATLAS:2012hna}. Some points are slightly shifted along the $x$ axis for better representation.}
\label{fig-v2npart}
\end{figure}

The path-length dependence of jet energy loss can be characterized by jet $v_{2}$, i.e., $v_{2}^{jet}$=$\left\langle cos2(\phi^{jet}-\Psi_{RP}) \right\rangle$, where $\phi^{jet}$ is the azimuthal angle of the jet and $\Psi_{RP}$ is the azimuthal angle of the reaction plane formed by the impact parameter $b$ and the beam direction which is fixed to $\Psi_{RP}$=0 in our AMPT simulations. Fig.~\ref{fig-v2npart} (a)-(d) show the comparison of $v_{2}^{jet}$ as functions of the number of participant nucleons ($N_{part}$) between the AMPT results and the ATLAS experimental data for different jet $p_{T}$ bins in Pb+Pb collisions at $\sqrt{s_{_{\rm NN}}}$ = 2.76 TeV. The AMPT results give qualitative trends similar to the experimental data, but slightly overestimate the magnitudes.

\begin{figure}
\includegraphics[scale=0.45]{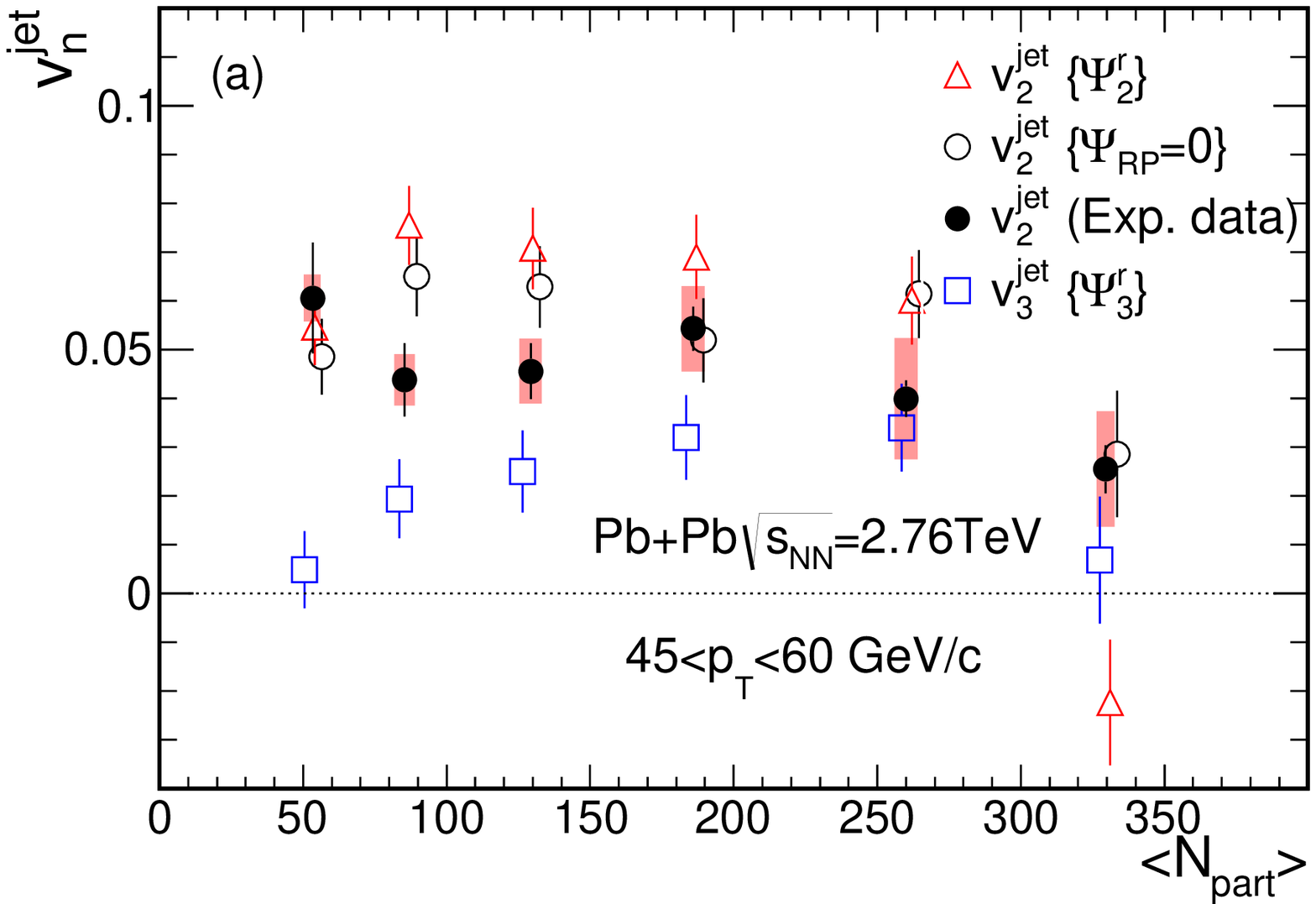}
\includegraphics[scale=0.45]{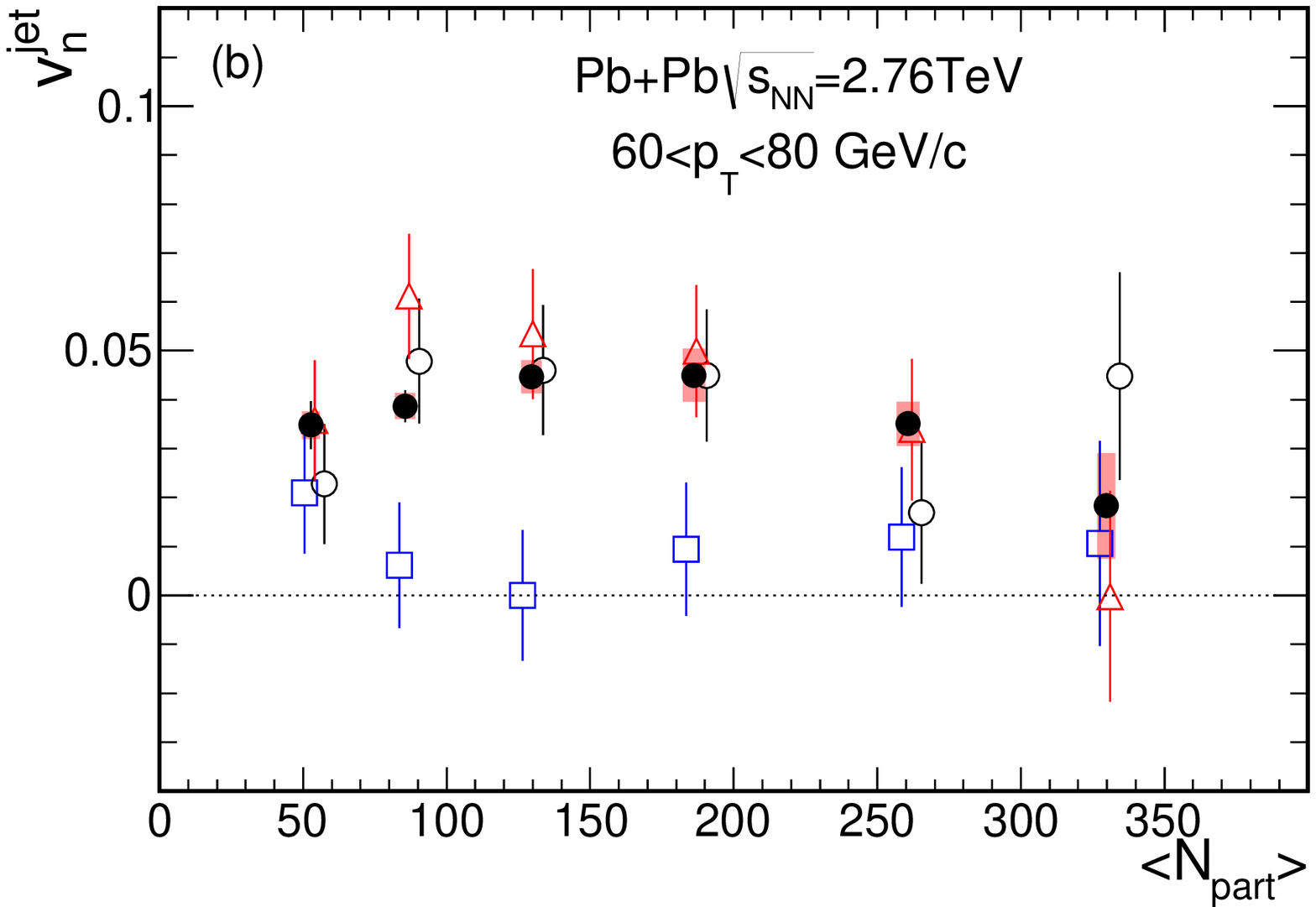}
\caption{(Color online) $v_{n}^{jet}$ (n= 2 and 3) as functions of $N_{part}$ for jet $p_{T}$ bins of $45< p_{T} <60$ GeV/$c$ (a) and  $60< p_{T} <80$ GeV/$c$ (b) in Pb+Pb collisions at $\sqrt{s_{_{\rm NN}}}$ = 2.76 TeV, where open triangles represent $v_{2}^{jet}$ with respect to $\Psi_{2}^{r}$, open circles represent $v_{2}^{jet}$ with respect to $\Psi_{RP}$ = 0, open squares represent $v_{3}^{jet}$ with respect to $\Psi_{3}^{r}$ and solid circles represent the ATLAS experimental data~\cite{Aad:2013sla,ATLAS:2012hna}. Some points are slightly shifted along the $x$ axis for better representation.}
 \label{fig-vnjet}
\end{figure}

It is well known that the odd orders of harmonic flows can arise from the initial geometry fluctuations through final state interactions~\cite{Alver:2010gr}. On the other hand, the even orders of harmonic flows are also affected if considering  the initial geometry fluctuations~\cite{Han:2011iy}. To calculate the n-th Fourier coefficient $v_{n}$, the n-th event plane $\Psi_{n}^{r}$ can be defined as
\begin{equation} \label{psi}
\Psi_{n}^{r}=\frac{1}{n}\left [arctan\frac{\left\langle r^{n}sin(n\varphi)\right\rangle} {\left\langle r^{n}cos(n\varphi)\right\rangle}+\pi \right],
\end{equation}
where  $r$ and $\varphi$ are the coordinate position and azimuthal angle of each parton in the AMPT initial state and the average $\langle \cdots\rangle$ denotes density weighting. Then the n-th harmonic coefficient of jets, $v_{n}^{jet}$, can be obtained by the following equation:
\begin{equation} \label{vn}
v_{n}^{jet}=\left\langle cos \left[ n(\phi^{jet}-\Psi_{n}^{r}) \right] \right\rangle.
\end{equation}
Note that the $v_{n}^{jet}$ definition is the same as that for a single hadron; however, $v_{n}^{jet}$ is expected to have smaller bias because the reconstructed jet has kinematic properties that are more closely related to those of the parent partons~\cite{Aad:2013sla,ATLAS:2012hna}.

Jet $v_{2}$ and $v_{3}$ as functions of $N_{part}$ for two typical $p_{T}$ bins of $45< p_{T} <60$ GeV/$c$ and  $60< p_{T} <80$ GeV/$c$, calculated by Eqs.~(\ref{psi}) and (\ref{vn}) and denoted as $v_{2}^{jet}\{\Psi_{2}^{r}\}$ and $v_{3}^{jet}\{\Psi_{3}^{r}\}$, are shown in Figs.~\ref{fig-vnjet} (a) and (b), respectively. $v_{2}^{jet}\{\Psi_{2}^{r}\}$ (open triangles) is consistent with the previous jet $v_{2}$ calculations of $v_{2}^{jet}\{\Psi_{RP}=0\}$ (open circles), though it has a little higher magnitudes due to the initial fluctuation contribution~\cite{Han:2011iy}. For jet $v_{3}$, it is smaller than jet $v_{2}$. By comparing jet $v_{3}$ between two different jet $p_{T}$ bins, jet $v_{3}$ tends to vanish with  increasing jet $p_{T}$.

\begin{figure}
\includegraphics[scale=0.45]{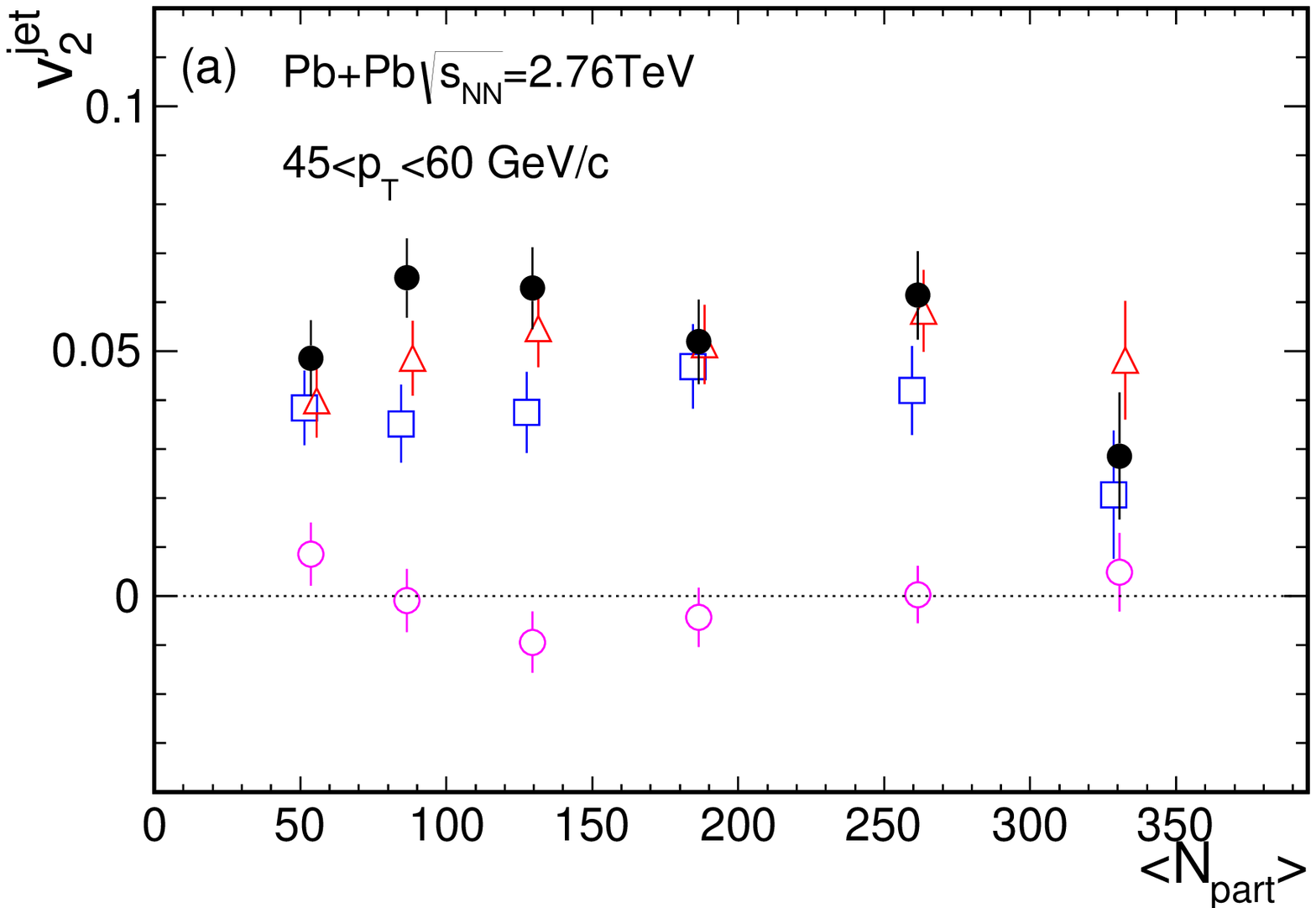}
\includegraphics[scale=0.45]{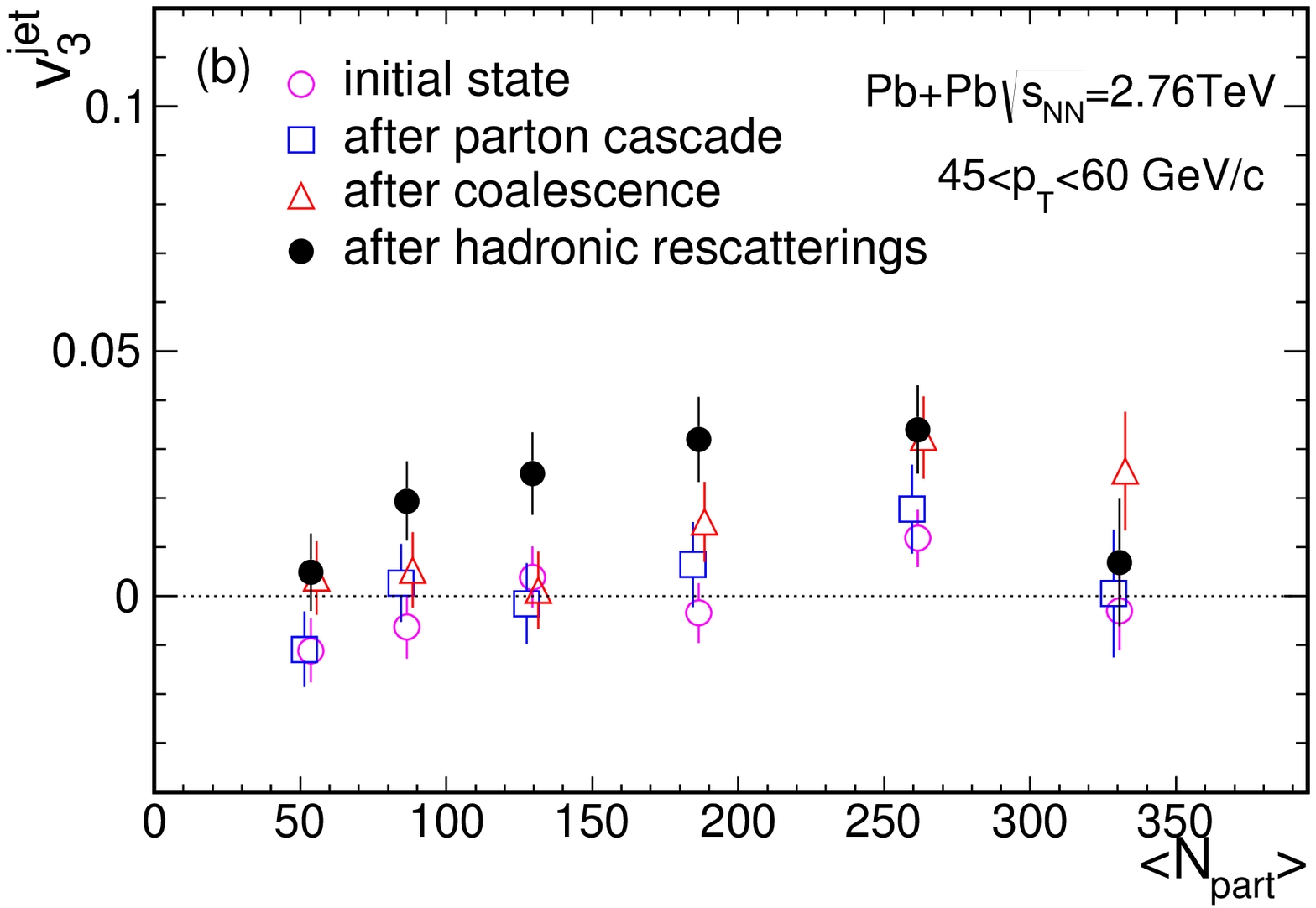}
\caption{(Color online) The AMPT results on $v_{2}^{jet}$ (a) and $v_{3}^{jet}$ (b) as functions of $N_{part}$ for the jet $p_{T}$ bin of $45< p_{T} <60$ GeV/$c$ for different evolution stages in Pb+Pb collisions at $\sqrt{s_{_{\rm NN}}}$ = 2.76 TeV. Some points are slightly shifted along the $x$ axis for better representation.
}
\label{fig-time}
\end{figure}

Since heavy-ion collisions are dynamical evolutions which involve many important evolution stages, it is important to investigate the stage evolution of jet $v_{n}$. Figs.~\ref{fig-time} (a) and (b) display jet $v_{2}$ and $v_{3}$ for the $p_{T}$ bin of $45< p_{T} <60$ GeV/$c$  at different evolution stages in Pb+Pb collisions from the AMPT simulations, respectively. The jet $v_{n}$ is nearly zero in the initial state.  However, jet $v_{n}$ arises from the process of parton cascade, which indicates jet $v_{n}$ is generated owing to the strong interactions between jet and the partonic medium. On the other hand, the processes of hadronization via coalescence and final hadronic rescatterings have little impact on jet $v_{n}$.

\begin{figure}
\includegraphics[scale=0.45]{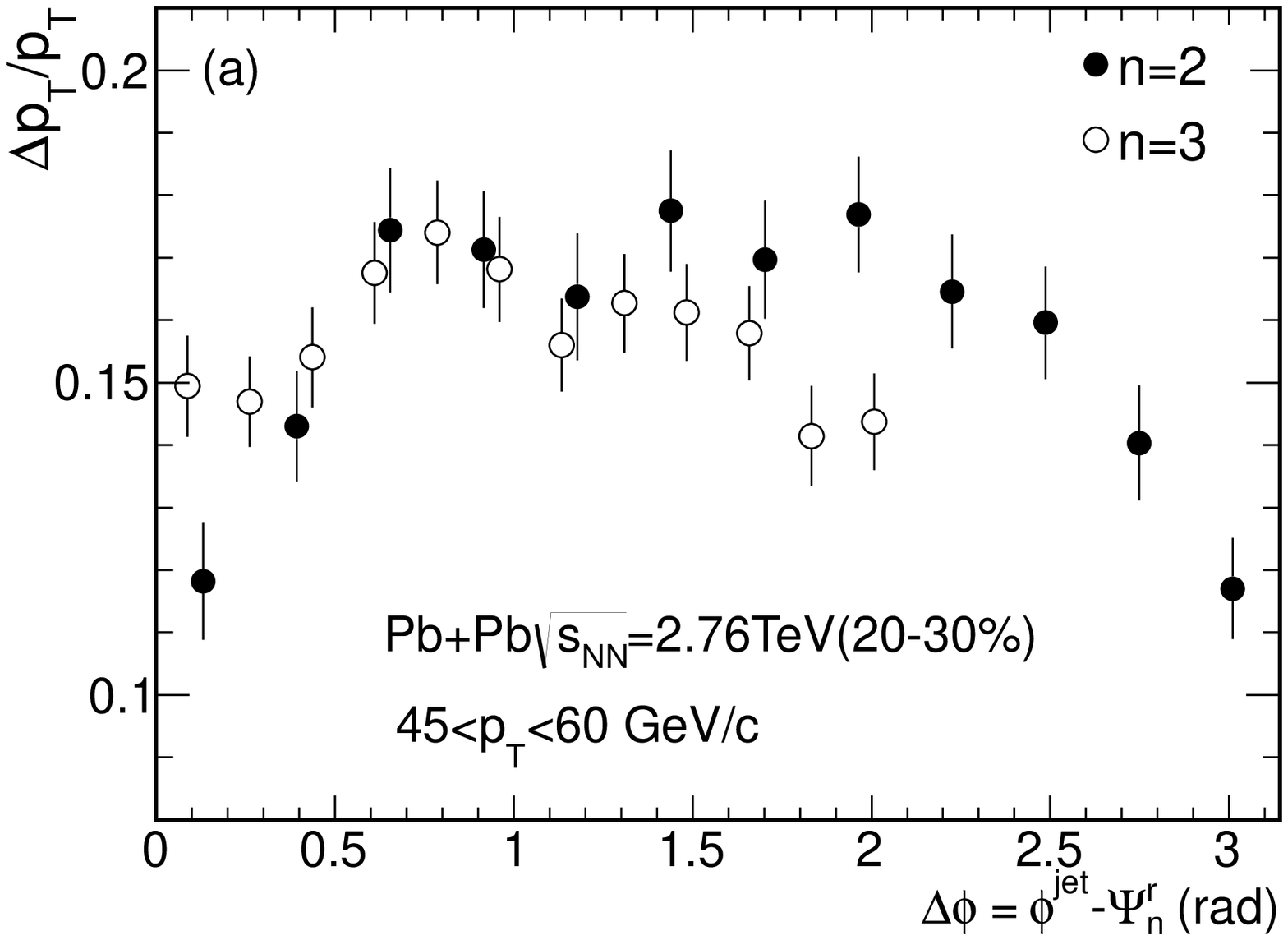}
\includegraphics[scale=0.45]{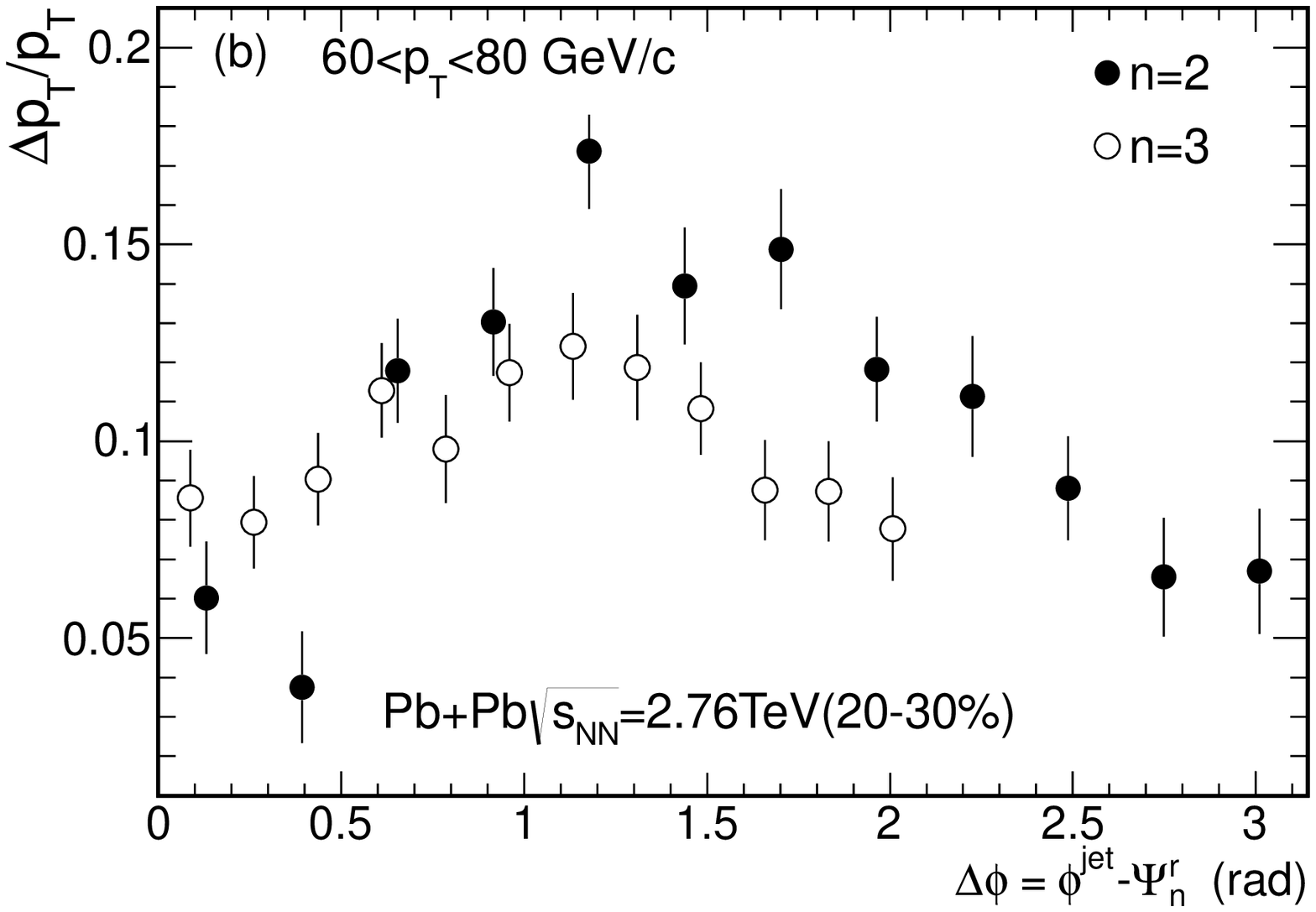}
\caption{The AMPT results on the jet energy loss fraction, $\Delta p_{T}/p{_{T}}$, as functions of $\Delta\phi$ = $\phi^{jet}-\Psi_{n}^{r}$ [n=2 (solid circles) and 3 (open circles)] for the jet $p_{T}$ bins of  $45< p_{T} <60$ GeV/$c$ (a) and  $60< p_{T} <80$ GeV/$c$ (b) in the centrality bin of 20-30\% in Pb+Pb collisions at $\sqrt{s_{_{\rm NN}}}$=2.76 TeV.}
\label{fig-Dpt}
\end{figure}

Figs.~\ref{fig-Dpt} (a) and (b) show the AMPT results for the averaged jet energy loss fraction $\Delta p_{T}/p_{T}$=$(p_{T}^{jet, initial}-p_{T}^{jet final})/p_{T}^{jet, initial}$ as functions of the relative azimuthal angle $\Delta \phi=\phi^{jet}-\Psi_{n}^{r}$ for two jet $p_{T}$ bins of $45< p_{T} <60$ GeV/$c$ and  $60< p_{T} <80$ GeV/$c$ in their first azimuth periods for the centrality bin of 20-30\% in Pb+Pb  collisions at $\sqrt{s_{_{\rm NN}}}$=2.76 TeV, respectively. Jets lose more energy at $\Delta\phi \sim \pi/2$ with respect to the second order of the event plane or $\Delta\phi \sim \pi/3$ with respect to the third order of the event plane. It can be reasonably understood because jets transverse a longer path length through the medium in the direction of $\Delta\phi \sim \pi/2$ or $\Delta\phi \sim \pi/3$ for an elliptic or triangle shape profile, which is consistent with the path-length effect of jet energy loss~\cite{oai:arXiv.org:nucl-th/0012092}.

\begin{figure}
\includegraphics[scale=0.45]{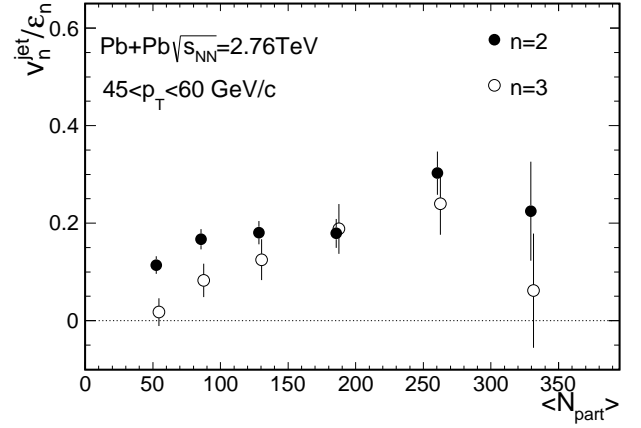}
\caption{ The AMPT results for $v_{n}^{jet}/\varepsilon_{n}$ [n=2 (solid circles) and 3 (open circles)] as functions of $N_{part}$ for the jet $p_{T}$ bin of  $45< p_{T} <60$ GeV/$c$ in Pb+Pb collisions at $\sqrt{s_{_{\rm NN}}}$=2.76 TeV. Some points are slightly shifted along the $x$ axis for better representation.
}
\label{fig-vntoen}
\end{figure}

\begin{figure}
\includegraphics[scale=0.45]{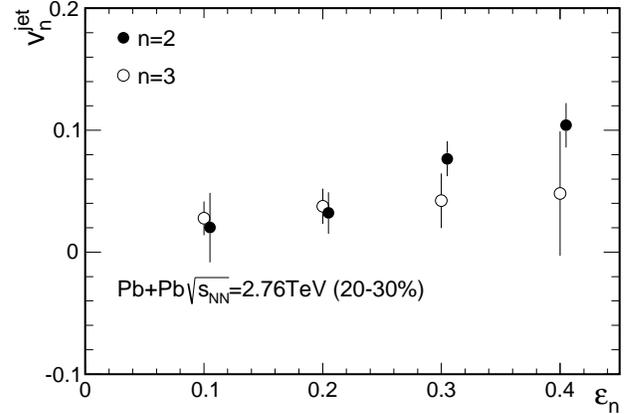}
\caption{ The AMPT results for $v_{n}^{jet}$ as functions of $\varepsilon_{n}$ [n=2 (solid circles) and 3 (open circles)] for the jet $p_{T}$ bin of  $45< p_{T} <60$ GeV/$c$ in the centrality bin of 20-30\% in Pb+Pb collisions at $\sqrt{s_{_{\rm NN}}}$=2.76 TeV. Some points are slightly shifted along the $x$ axis for better representation.
}
\label{fig-vnepsilon}
\end{figure}

The conversion efficiency ($v_{n}/\varepsilon_{n}$) has been used as an important observable to understand the collective flow phenomena in high-energy heavy-ion collisions~\cite{Han:2011iy,Alver:2010gr}. Similarly, $v_{n}^{jet}/\varepsilon_{n}$ could disclose how jet quenching depends on the initial geometry shape. To calculate the n-th order eccentricity $\varepsilon_{n}$, we use the definition as follows,
\begin{equation} \label{ep}
\varepsilon_{n}=\frac{\sqrt{{\left\langle r^{n}sin(n\varphi)\right\rangle}^{2}+{\left\langle r^{n}cos(n\varphi) \right\rangle}^{2}}}{\left\langle r^{n} \right\rangle},
\end{equation}
according to the information about the coordinate space of initial partons. Fig.~\ref{fig-vntoen} shows the AMPT results for $v_{n}^{jet}/\varepsilon_{n}$ as functions of $N_{part}$ for the jet $p_{T}$ bin of  $45< p_{T} <60$ GeV/$c$ in Pb+Pb collisions at $\sqrt{s_{_{\rm NN}}}$=2.76 TeV. The $v_{n}^{jet}/\varepsilon_{n}$ increases with $N_{part}$ except in the most central centrality bin where jet $v_{n}$ is close zero, which reveals that azimuthal anisotropies of jets are more easily formed in more central collisions owing to the larger jet energy loss in a denser partonic matter.  Fig.~\ref{fig-vnepsilon} presents jet $v_{n}$ as functions of the eccentricity $\varepsilon_{n}$ for the jet $p_{T}$ bin of  $45< p_{T} <60$ GeV/$c$ in a selected centrality bin of 20-30\% in Pb+Pb collisions. It is shown that the final jet $v_{n}$ increases with the initial spatial eccentricity or triangularity, which indicates that jet azimuthal anisotropies are produced by the interactions between jets and the partonic medium with different asymmetrical geometry shapes. 

\section{CONCLUSION}
\label{sec:conclusion}

In conclusion, azimuthal anisotropies of reconstructed jets have been investigated in Pb+Pb collisions at $\sqrt{s_{_{\rm NN}}}$=2.76 TeV within a framework of a multiphase transport (AMPT) model. The model gives a qualitative description about the measured $v_{2}$ of reconstructed jets for the $p_{T}$ range from 45 to 160 GeV/$c$. We predict that $v_{3}$ of reconstructed jets, which has a smaller magnitude than its $v_{2}$, approaches zero with increasing jet $p_{T}$. It can be attributed to the dependence of the jet energy loss fraction on the azimuthal angles with respect to the different orders of event planes. The dynamical stage evolution of reconstructed jets discloses that jet $v_{n}$ mostly arises from a strong parton cascade process with little effect from the final stages such as hadronization and hadronic rescatterings. The ratio $v_{n}^{jet}/\varepsilon_{n}$ increases with $N_{part}$ in non-central Pb+Pb collisions, furthermore, jet $v_{n}$ increases with the initial spatial asymmetry ($\varepsilon_{n}$) for a given centrality bin. These behaviors indicate that jet $v_{n}$ is produced by the strong interactions between the jet and the partonic medium with different initial asymmetrical geometry shapes. Therefore, the azimuthal anisotropies of reconstructed jets can be utilized as a good probe to study the initial spatial asymmetry, and imposes constraints on the path-length dependence of jet quenching models.

\section*{ACKNOWLEDGMENTS}

This work was supported by the Major State Basic Research Development Program in China under Contract No. 2014CB845404, the NSFC of China under Projects No. 11175232, No. 11035009, and No. 11375251, the Knowledge Innovation Program of CAS under Grant No. KJCX2-EW-N01, the Youth Innovation Promotion Association of CAS, the project sponsored by SRF for ROCS, SEM, the CCNU-QLPL Innovation Fund under Grant No. QLPL2011P01, and the ``Shanghai Pujiang Program" under Grant No. 13PJ1410600.

\end{document}